# Perceptual Effects of Hierarchy in Art Historical Social Networks

Houda Lamqaddam, Inez De Prekel, Koenraad Brosens, Katrien Verbert

## 1. Introduction

Networks are some of the most commonly found visualizations in digital humanities research. Used to represent relationships between persons or concepts, this type of graphs has also been implemented through various layouts to better support specific tasks and users.

The wide availability of layout algorithms has multiple advantages for researchers who aim to analyze social networks. Force-directed layouts such as the Gephi Atlas drastically simplify the visualization – therefore analysis – of large network data. They reveal patterns and clusters by using complex physical models, all with limited need for user intervention or mathematical expertise. These models excel in domains where datasets are large, and where the goal is to identify the structure – or shape – of the dataset.

However, the widened access to these automated graph drawing algorithms drives us to interrogate their fitedness for the analysis of humanistic material. On the one hand, force-directed layouts visually remove existing structures in the data, even when these structures are meaningful to users and their daily tasks. On the other, humanistic datasets are often small, but do consist of meaningful inherent clusters, distances and relationships. Such a mismatch leads to representations that can be at odds with users' internal perceptions of the data structure. For instance, family members in a correspondence graph are likely to be spread apart. Similarly, non-contemporary actors can be placed side to side if they share a common correspondent. This inconsistency between mental models and visual representation can become a challenge for users. More critically, it has the potential to impact the adoption of visualization tools as a whole as research suggests usability and user trust, rather than scalability, are main obstacles to adoption of digital tools in the humanities [Thoden, 2017].

In this paper, we address this challenge by introducing *Force-Layered graphs*: a novel method for drawing hierarchical graphs[1] using force-directed layout. This method uses the existing – and already mature - dynamics of force-directed algorithms, and injects a meaningful structure into them to organize them in a way that better supports users' mental models. We describe the results

[1] also known as hierarchical or Sugiyama-style graphs.

of a mixed-methods user study exploring the effect of visual hierarchy for users with humanities background.

Our initial results reveal that force-layered graph layout outperforms force-directed layout in terms of number and depth of insights, cognitive load reduction, and user preference. We discuss these findings along with implications for design and future directions for this line of research. Specifically, we interrogate the artificial structure these layouts create, and evaluate whether their disconnect with existing data hierarchies affects user understanding, preference, and experience.

# 2. Structure & Hierarchy in Network Visualization

Discussion about the semantics of visual design elements is often neglected within technical scholarship on information visualization. In the case of hierarchy, we believe this discussion to be critical for two reasons. First, because Human-Computer-Interaction research has shown the importance of the consistency between users' mental models of information and its digital representation [Dørum, 2011]. Second, because visual hierarchy in network graphs strongly affects their spatial organization, and spatial position is believed to be one of the most effective channels in information visualization [Munzner, 2014; Bertin, 1967]. In this section, we present a discussion on the semantics of hierarchical trees on one side, and force-directed networks on the other.

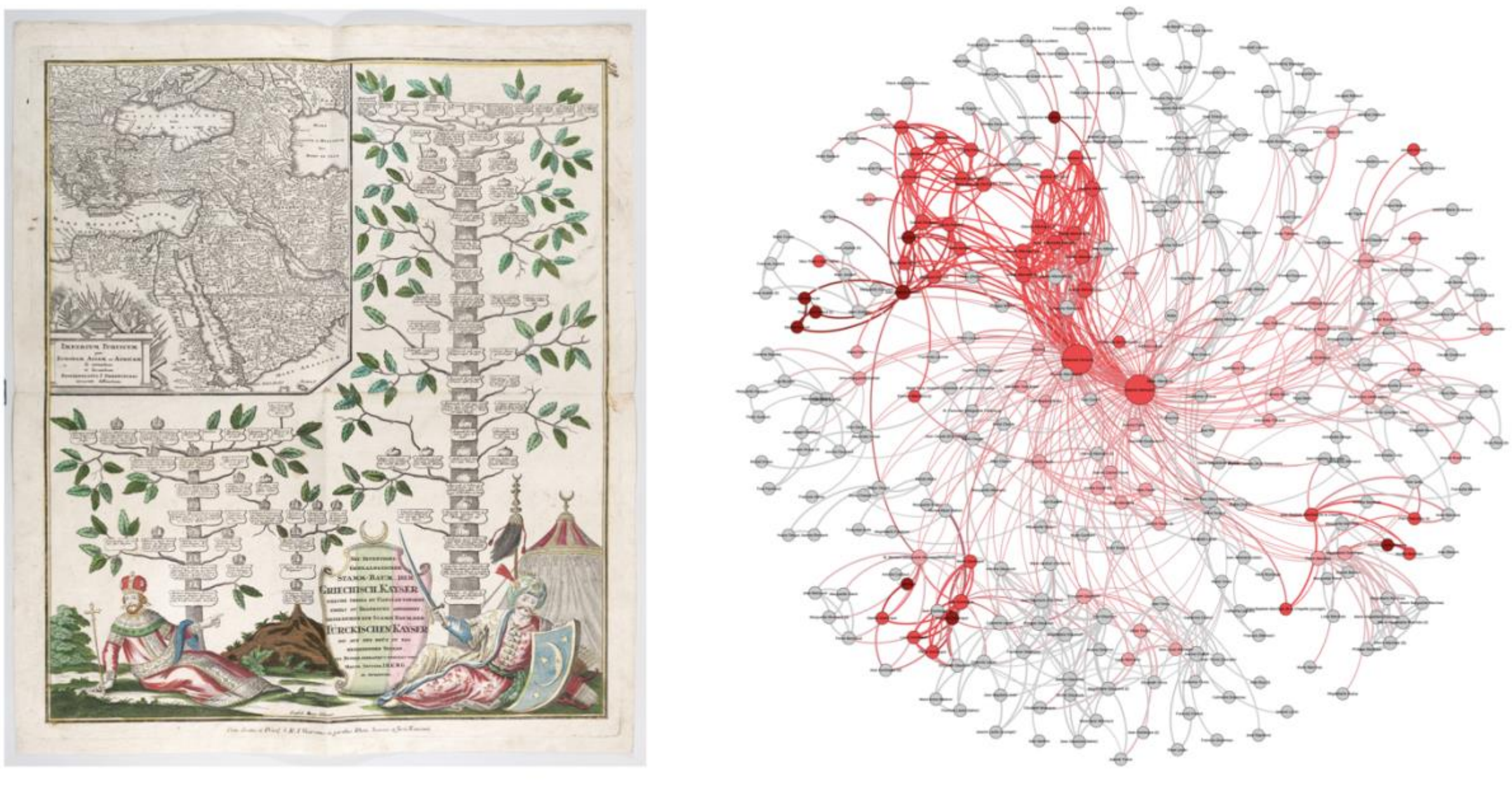

*Figure 1 (a) On the left: "A newly invented genealogical tree of the last Byzantine emperor...as well as one of the Turkish sultan". ca. 1730-57. Digital image from the Met Museum Collection Gift of Robert W. Hatem. Public Domain. (b) On the right: a network showing relationships between signatories of a 18th century marriage contract in France. This network is an example of graph drawn using a circular[2] force-directed layout. [Rothschild, 2014]. Reproduced with permission.*

[2] This layout is a specific type of force-directed layout that confines nodes to a circular space. The Fruchterman-Reingold layout on Gephi is an example of such algorithms.

First, we illustrate the difference between the two, through the example of two visualisations of historical social networks as can be seen in Figure 1. In Figure 1a, a hand-drawn illustration displays the genealogy of the last Byzantine emperor drawn as a collection of trees. This visualization has a strong spatial organization based on family lines, roles, and generational hierarchy. In fact, tree structures can be found in multiple historical representations of networks because they mirror the conception of lineage as a branching linear connected structure. In Figure 1b, we see a contemporary representation of a historical network[3], this time depicting the list of signatories of a marriage contract in Angoulême, France in the 18th century, as well as their extended social networks [Rothschild, 2014]. This network, developed using Gephi shows the global social structure of this historical community, produced using a force-directed layout, such that highly connected subgroups attract each other, thereby revealing the cliques and subcommunities within the dataset.

To describe them formally, hierarchical trees (such as Figure 1a) are a sub-category of networks that are acyclic (i.e. there are no loop encountered during a traversal of the tree) and directed (i.e. the relationships described by the edges are one-directional, rather than reciprocal). The hierarchical aspect signifies a certain categorization of the nodes so that those placed on the same level hold a similar value or meaning. Hierarchical trees are a spatial organization whose traces date back to antiquity, and which can be found in a variety of knowledge domains. As described by Drucker in her book Graphesis [Drucker, 2010], trees are interesting because their spatial organization carries meaning. In the case of social networks, this spatial organization becomes an indicator of multiple elements including birth order, generational breadth and span, patterns of marriage [Drucker, 2010]. Another interesting aspect to hierarchical trees is their embedded story-telling. The root node(s), typically placed at the top[4] of the hierarchy, creates a visual anchor that drives the reader in, and directs their attention down the edges in a manner which is almost a narrative approach to visualization.

On the other hand, force-directed networks (Figure 1b) are mathematical products, where physical simulations ran multiple times on a dataset model attraction and repulsion between nodes, therefore automatically placing nodes with high connectivity close to one another, or at the opposite, resulting in disconnected nodes being placed far from each other, highlighting their disconnect. For Ahnert et al., force-directed networks "reject hierarchies", which become hidden and difficult to notice, even if they were strongly present within the dataset [Ahnert, 2020]. Modern readings have argued for the advantages of force-directed networks by justly questioning the relevance and accuracy of representing the world in rigidly structured hierarchical roles. In actual historical communities, influence does not necessarily traverse time and generations in a clear-cut linear way. Similarly, trees of knowledge fall short when representing complex webs of information, as the distributed aspect of the web and social networks have clearly illustrated in the past decades[5].

[3] We defaulted to a contemporary representation for the non-hierarchical layout as these can only be found in recent representations. See [Freeman, 2004] for a history of network representation of social structures.
[4] Or at the right, in a right-to-left horizontal spatial organization

In the context of humanistic research, however, the intentional organization of network data has an undeniable value, when compared to the visual organization that results of mathematical algorithms. Drucker explains it clearly:

> *"The spatial distribution of network diagrams, topic maps, and other graphical expressions of processed text or intellectual content is often determined by the exigencies of screen real estate, rather than by a semantic value inherent in the visualization. This introduces incidental artifacts of visual information. A point in a graph may be far from another because of a parameter in the program that governs display, rather than on account of the weight accorded to the information in the data set. The argument of the graphic may even be counter to the argument of the information, creating an interpretative warp or skew, so that what we see and read is actually a reification of misinformation."* [Drucker, 2014]

In that way, network visualizations are no different than other visualizations of data, as they themselves can reproduce existing biases, and cement them as neutral knowledge. Porras asks us to keep our eyes open to the dangers of network visualizations as they can "reinscribe historic and contemporary power differentials" [Porras, 2017].

In visualization literature, a growing voice argues for the consideration of semantics and connotation in the visualization design, especially within humanistic scholarship. In previous work [Lamqaddam, 2020], we report how humanists have consistently commented on the connotative dryness of visualization tools, linking it to a reduction of rich research material [Manovich, 2011]. Through user research, we found that humanist scholars felt a *semantic distance* between their experience of their research material and practice and its representation in visualization tools. Such tools can end up stripping data of essential meaning-making elements such as temporality, physicality, and terminology. More specifically, we identified structure in conceptualizations and ontologies as one of the characteristics of humanistic research that is often missing in visualizations. We also found proof that meaning, intentionally layered into visualization, had the potential to reduce the semantic distance experienced by scholars, and alleviate their discomfort. In this context, the question of representing data's hierarchical structures within historical social networks is an interrogation of our Layers of Meaning (LoM) framework, and an investigation of a larger trend in digital humanities scholarship during the data era.

---

5 Pushed to its extreme, this urge to categorize the world can be linked to historical movements which have led to similar attempts at categorizing peoples and communities, thereby creating and intensifying the construction of race for example, solidified through its representation as formal scientific truth. This discussion, while outside the scope of this article, cannot be ignored when discussing the role of hierarchy and knowledge in human history.

# 3. Research Questions

To build on the previous discussion, we wanted to investigate the effects that hierarchical trees and force-directed graphs have on user perception. Because the literature on mental models in visualization shows link effects on efficiency and user experience, we developed an extensive user evaluation protocol where we compare the two representations and investigate the perceptual impacts of each one. In the next section, we describe the process we followed, and the findings we can extract from it.

Through this process, we aimed to answer the following research questions:

- RQ1: Do users have an inherent mental model of social network spatial organizations? How is it visually structured?
- RQ2: How does representing historical social network data using a generational hierarchy affect cognitive load during social networks analysis tasks?
- RQ3: How does hierarchy within social network data affect insight-building and recall?

In order to perform a fair comparison, we developed two visualizations showing both hierarchical and standard force-directed visualizations of historical social networks. We developed a prototype for hierarchical tree layout visualization using the Javascript visualization library D3. We chose to develop this algorithm on top of the existing force-directed layout algorithms for multiple reasons. First, the existing force-directed layout algorithms are currently mature and deliver high-quality performance. Perhaps as a consequence, these layouts are supported by major graph drawing libraries and toolkits, which opens future possibilities for wider implementation of hierarchical graph layouts. Force-directed layouts also include parameters to reduce node collision, occlusion, and bring in useful results, such as bringing together highly connected clusters, which, while we do not want it to break inherent hierarchy in the data, still proves to be useful in reducing edge-crossing.

We designed and implemented an algorithm that traverses a force-directed graph and spatially organizes it based on a user-defined parameter. The full algorithm description and code can be found on our GitHub page[6]. Figure 2 shows the same dataset represented as both a force-directed graph (left), and a hierarchical network (right). We call the graphs produced by our hierarchical layout *force-layered graphs.* This porte-manteau hints at the underlying force-directed layout, and includes the layered aspect of hierarchical graphs (alternatively referred to as layered graphs, or Sugiyama-graphs).

[6] github.com/HLMQA/egoNahr

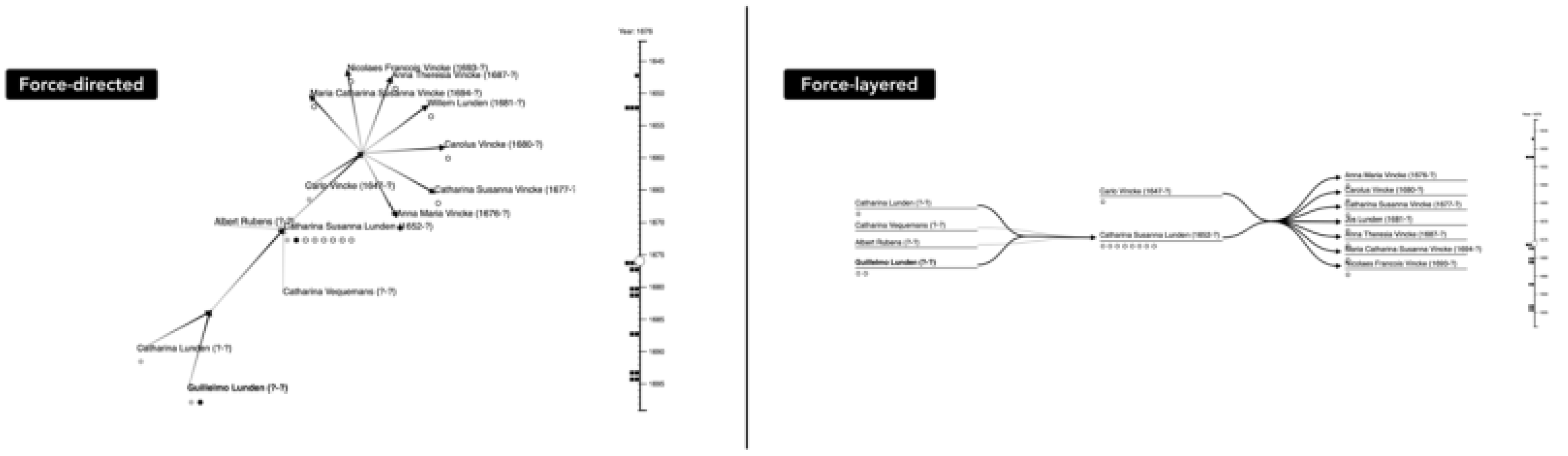

*Figure 2 A dataset of thirteen nodes is visualized using a force-directed layout (left) and a force-layered layout (right).*

# 4. Study

In this section, we describe the evaluation protocol used to answer the RQs described in section 3. In HCI, user evaluations are a standard way to get feedback and to get better insight into users' needs. It is also a common way to evaluate usability and identify obstacles to adoption of digital tools. Since usability and adoption are a recurrent issue in DH research, there is real value to be gained from applying such methods to pinpoint exact usability issues in collaboration with the intended audience. In this section, we describe each step of the methodology together with the results found. For each subsection, the full table of results can be found in the annex. In this section, we sometimes use the notations *F-D* and *F-L* as shorthand for *Force-Directed* or *Force-Layered* graphs respectively.

***Participants.*** We recruited 15 participants (12 women, 3 men) with background in Art History and Digital Humanities, most of them graduate students in the university's Digital Humanities MSc program. Participants had on average 3.6 years of experience in their domain, and 1.2 years of experience using digital tools to support their research or study practice. Among participants, most digital tool usage within field of study consisted of digital archives (11 out of 15), spreadsheets (10 out of 15), and graph visualization tools (9 out of 15).

***Data.*** We used data extracted from the Cornelia[7] database, which uses archival documents to gather historical data on 17th century artistic communities[8]. We selected a subset containing 38 persons related through ancestry, marriage and godparenthood links. This dataset specifically highlights (extra)familial hierarchy structure.

***Graphs.*** Figure 3 shows the two network visualization used in this study. We used the same dataset for both graphs not to introduce bias. However, to minimize learning effects, we manually changed the names so that the similarity of the two networks would not be obvious.

[7] Projectcornelia.be
[8] Specifically Flemish painters and tapestry producers.

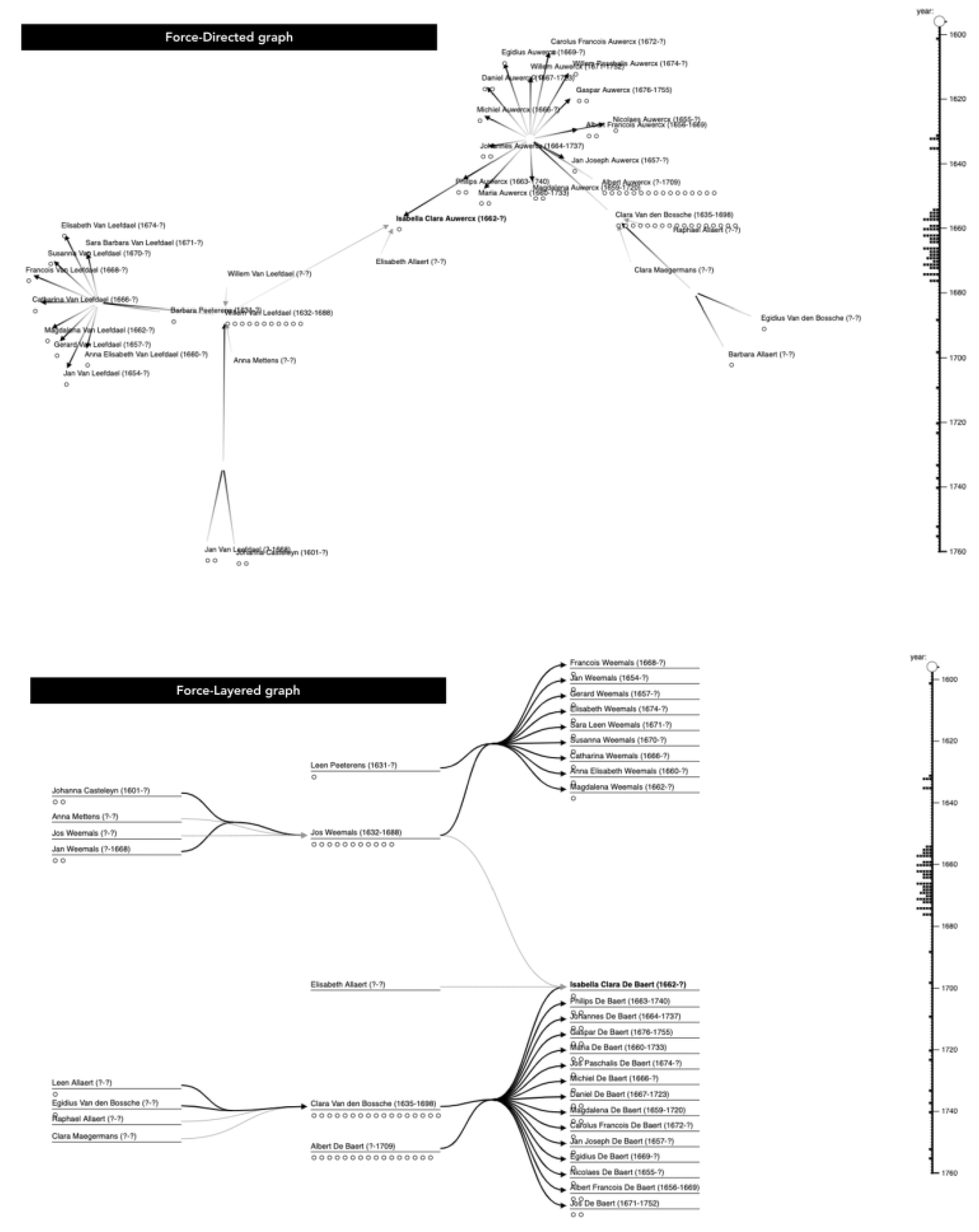

*Figure 3 The prototypes used in the user study. (a) On the top, the D3 force-directed layout. (b) On the bottom, the same data using our force-layered algorithm. The names of the actors were changed in both graphs to reduce learning effects on participants.*

# 4.0 Mental model elicitation

Understanding users' existing mental models (i.e. the representation they internally make of specific information structures or processes) is key if we aim to support similar structures in building visualizations. Previous work in the HCI field has demonstrated the value of asking participants to draw their mental models as a way to externalize them [Zhang, 2008; Kodama, 2017].

Before showing participants the graphs, we first started by asking each one to freely draw a social structure of their choosing, in order to elicit their mental mapping of social structures. They were invited to sketch out any fictional or existing social structure network (from stories, movies or books, or their own personal life[9]). They were encouraged to consider lineage relationships, but also non-blood ties such as extended families, and close friends. We then extracted the recurring themes present in the produced sketches. A few of these sketches can be seen in Figure 4.

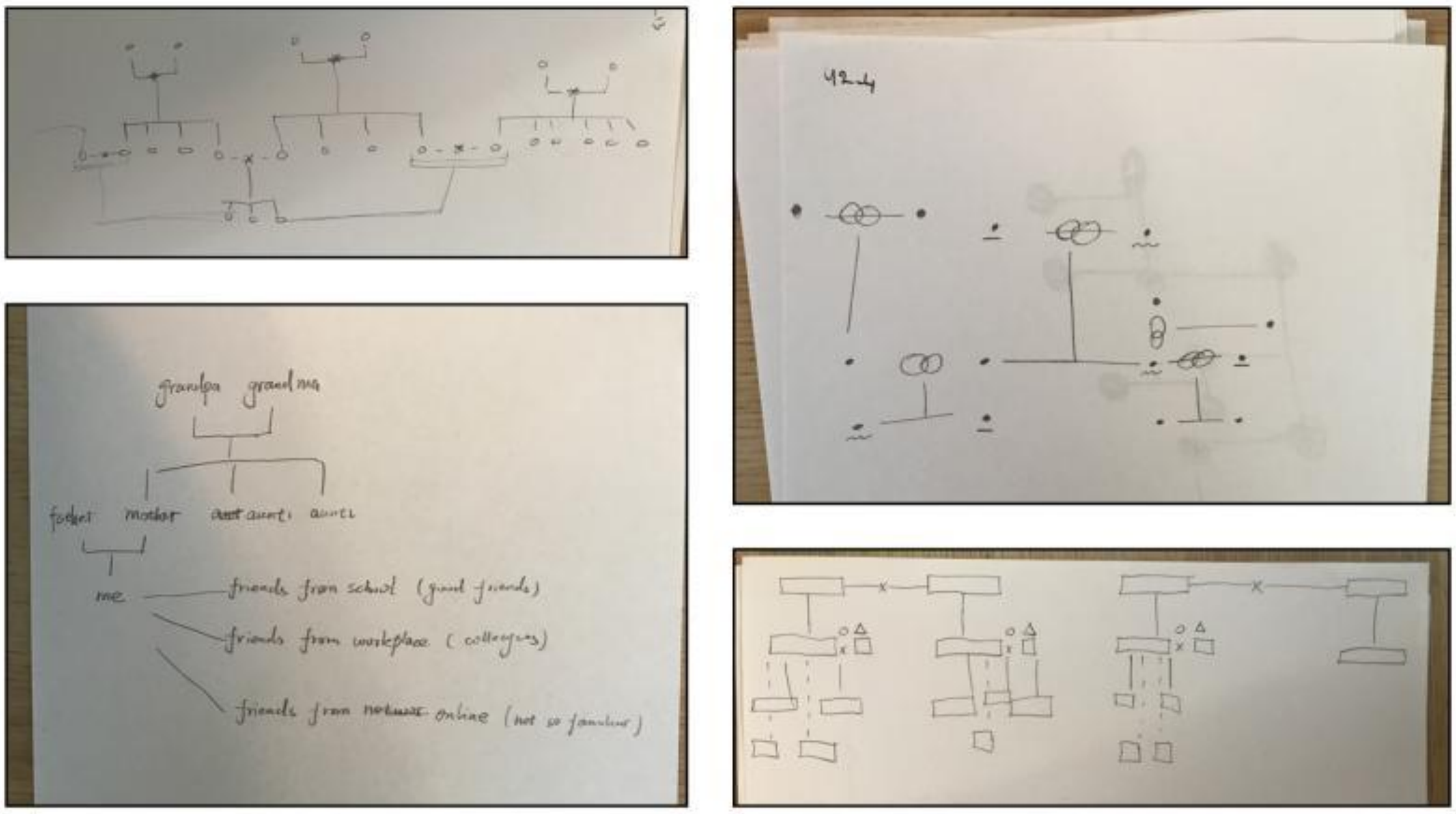

*Figure 4 From left to right, top to bottom: initial drawings for participants P13, P4, P15 and P10. These sketches show networks represented as layered graphs, similar to traditional ancestry tree visualizations. Participants made use of labels, symbols and shape.*

We found that in all[10] cases, participants sketched graphs had a hierarchical structure. These representations were mainly tree structures, with many of them structured as layered graphs. We also found that most participants used the vertical axis as a single indicator of generational division. However, in some cases, this pattern was broken as different relationships (edges towards friends for instance) were represented as orthogonal to the family links (Figure 4, P15). Indeed, almost a third (35\%) of participants chose to represent multiple relationship types within the graphs.

***Familiar structure, or expression of an internal mental model?*** While we can use the results from this elicitation task to induce the existence of a 'natural' hierarchical mental model of such data for the users, the exposure to family-tree type representations is an equally likely explanation for the phenomenon. Family trees are an old form of representations that are commonly seen by the public [Mitchell, 2014]. They are also commonly used in (art) historical teaching, which makes our participants even more likely to have built familiarity towards them. In reality, our participants have likely interacted with similar representations in the past, and may have defaulted to recreating a similar structure. We consider that these results do inform us of the internal model within the participants, itself caused by either inherent hierarchization of this data, or unvoluntary training caused by previous exposure.

[9] We asked participants to anonymize any personal information in the case they wanted to draw their personal social structure, in order to maintain privacy for them and their social circles.
[10] We excluded a single sketch from our analysis (P11) as it represented an evolutionary tree rather than a social network.

## 4.1 Think aloud study

We asked participants to explore the dataset in each condition while vocalizing their thought process. Similar to the mental model elicitation approach, this method allows to make the thought process of participants explicit. We then captured and transcribed participants think-aloud process and analyzed it using a thematic analysis to uncover patterns of insight[11] and identify differences between the force-directed and the force-layered conditions.

We found that the force-layered visualization triggered more interpretative[12] and reflective insights. We also find more occurrences of intrigue or curiosity about the portrayed social structures ("*It's too bad you can't see more about the person, you can't tell for example what their profession is", "I don't know the person, but I imagine if I were researching these people, it can be really interesting to see"*). Some participants also found contradictions in the underlying data, and offered hypotheses to explain those, or simply pointed them out ("[This person] *has an offspring but it's somehow not an event for* [their wife] -- *it's not an event for her, although it is with her"*). Finally, we found a clear divide in the number of negative emotions such as overwhelmingness or defeat in the condition of force-directed network usage (*"There are a lot of persons, and here a lot of events. I'm trying to see who's linked to who but it's really difficult here.", "I don't like this one. Yeah it's really confusing and I don't find … ugh* [trails off]*"*).

## 4.2 Exploration Tasks

To complement the open nature of the think-aloud exploration, we conducted a more structured exploration task where we asked participants to answer specific questions about the social dynamics portrayed in the visualization. We designed these questions based on the existing literature on graph exploration tasks[13] and categorized them into easy questions (e.g. "*find the most*

[11] In order to categorize the insight level of participants, we used the framework defined by Claes et al. [Claes, 2015] distinguishing three levels of insight depth: factual insight, referring to a description of visible data, interpretive insight, where data is synthesized along with additional objective or experienced information, then reflective, which also includes subjective or emotional response.
[12] In our study, factual insight describes statements such as ``he dies in 1668" or ``this person has children". Interpretative insight includes ``it's just close family I see", ``This person has children, but not with this one I think". Finally, in reflective insight, we included comments about the data quality (``I got the feeling this [dataset] is more documented than the other one") and emotional engagement with the data (``oh they had *many* children *laughs*").
[13] We used the taxonomy proposed by Lee et al. [Lee, 2006] and partly aimed towards evaluators who wish to compare graph visualizations. From a domain point of view, we looked at Lamqaddam et al. task categorization [Lamqaddam, 2018], which is more specific to the domain of family tree visualization in the humanities.

*connected actor*", "*Find all people who knew both X and Y*") and difficult questions ("*How does the graph evolve between birth of person X and their death*", *"Describe the network at a specific year"*).

We assessed the answers to the exploration tasks based on complete and partial correctness. We found that for easy questions, a similar number of participants (F-D: n=11; F-L: n=10) gave correct answers with both conditions. However, in the force-directed condition, half of these answers were incomplete. To the difficult questions, more participants gave correct answers using the force-layered condition (n=6). Notably, however, none of the participants were able to answer the same questions using the force-directed condition completely and correctly. The number of participants who gave incomplete answers was the same for both conditions.

## 4.3 Perception Task

In order to measure how the perception of the graphs compared with their actual characteristics, we collected participants recollection of some quantitative graph characteristics such as network size, and number of generations. We also asked about the perceived complexity of the social structure. Both networks contained the same number of nodes thirty-eight nodes. We found that participants perceived the force-layered graph to contain *fewer* nodes (median=30) than was the case. They were more slightly accurate when using the force-directed graph (median=35).

Finally, a majority of participants reported feeling that the social structure displayed in the force-layered condition was of a simpler nature one than the force-directed. This corroborates the difference in perceived graph size, as participants perceived networks in the force-layered condition to have fewer nodes.

## 4.4 Recall task

To assess recall, we aimed to see how well participants remembered the graph in its entirety. We then asked them to draw the graph to their best recollection after having diverted their attention for a brief moment[14]. We then analyzed the resulting sketch based on categories including the restitution of meaningful sub-communities, node-link structure, and visual detail.

We found differences in specific metrics between the recall sketches of the two versions, specifically on the identification of bridge nodes, as well as the retention of the multiple types of relationships. In the force-directed recall sketches, participants have retained the "dog-bone" structure of the graph consisting of two large clusters representing the two main families and tied together through a bridge node. Most participants also retained - at least partially - the lower antennas formed by the two loosely connected pairs nodes. In the force-layered condition, the drawings have retained the left-to-right organization of the nodes, as well as the two distinct subgraphs representing the two families. The bridge node bringing these two groups together was not systematically included. Finally, participants retained the existence of multiple types of relationships better in the force-layered condition, as they recalled more of the godparent type links. Example drawings from participants for each condition can be seen in Figure 5.

[14] We used this time to ask participants to fill out the NASA-TLX questionnaire, described in the next paragraph. This step had the advantage to distract participants from the screen and the graph itself, while not being too distracting that it led them to unrelated trains of thought.

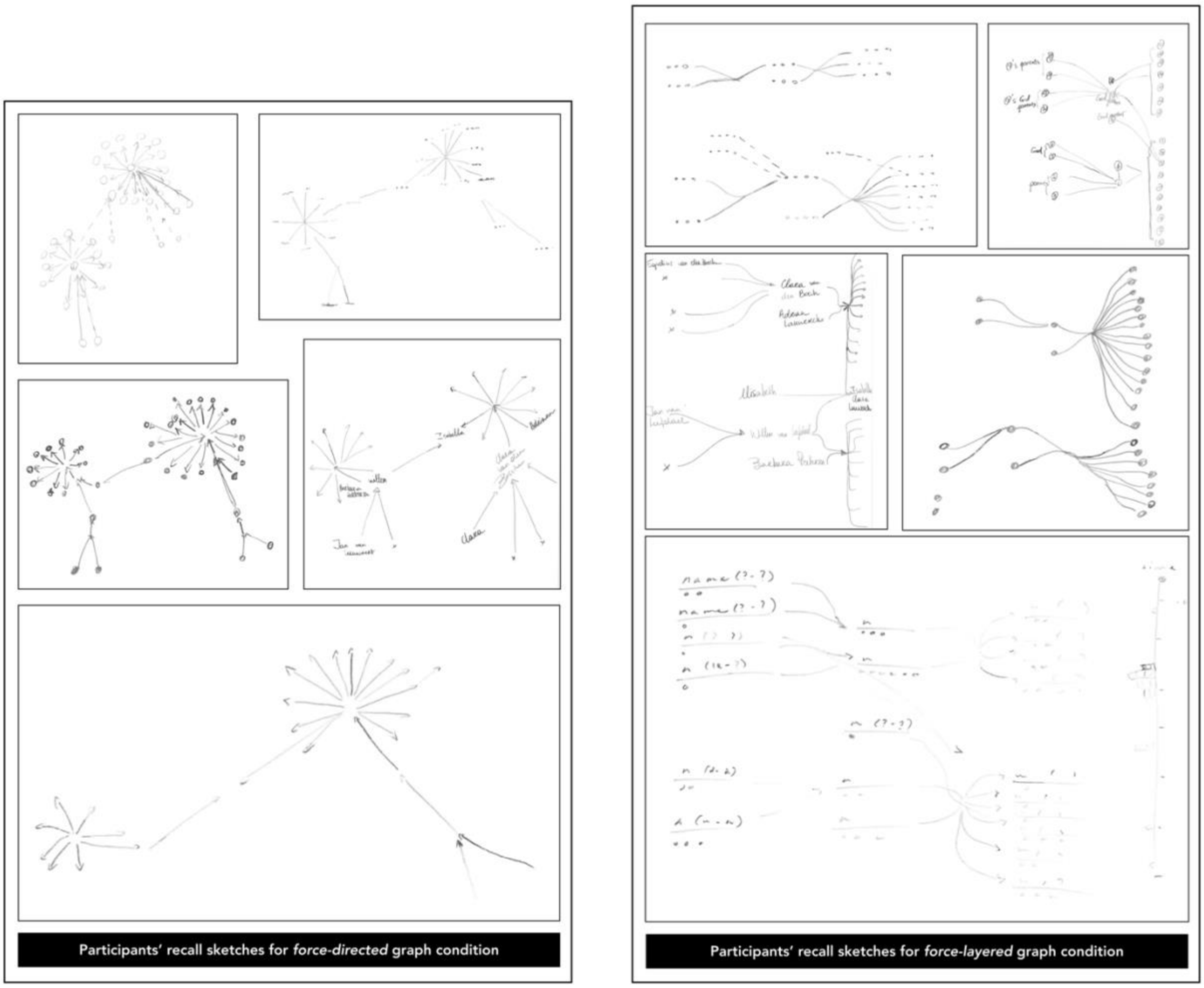

*Figure 5 Selected participants recall sketches in each condition. On the left, sketches representing the force-directed network (3a). On the right, sketches representing the force-layered (3b).*

## 4.5 Cognitive load

While evaluating performance can provide a good understanding of efficiency achieved using a system, the mental effort required is not answered by these approaches. We therefore measure participants' cognitive load[15] when using the two conditions.

As a measure for cognitive load, we used a raw NASA-TLX questionnaire [Hart, 1998] The NASA-TLX questionnaire is a widely used metric for measuring cognitive load in the HCI field. It consists of a self-reported scale capturing workload across six dimensions: Mental, Physical, and Temporal Demands, Frustration, Effort, and Performance [Hart, 1998]. Participants are asked to score each of these variables on a 0--100 scale where a lower value represents a lower demand[16].

Comparing the raw NASA-TLX scores, we find that the force-layered visualization (average score of 26.3) was found to be less demanding than the force-directed graph (average score of 35.78), the metrics with most variance between conditions being mental demand, effort, and frustration (see Figure 6).

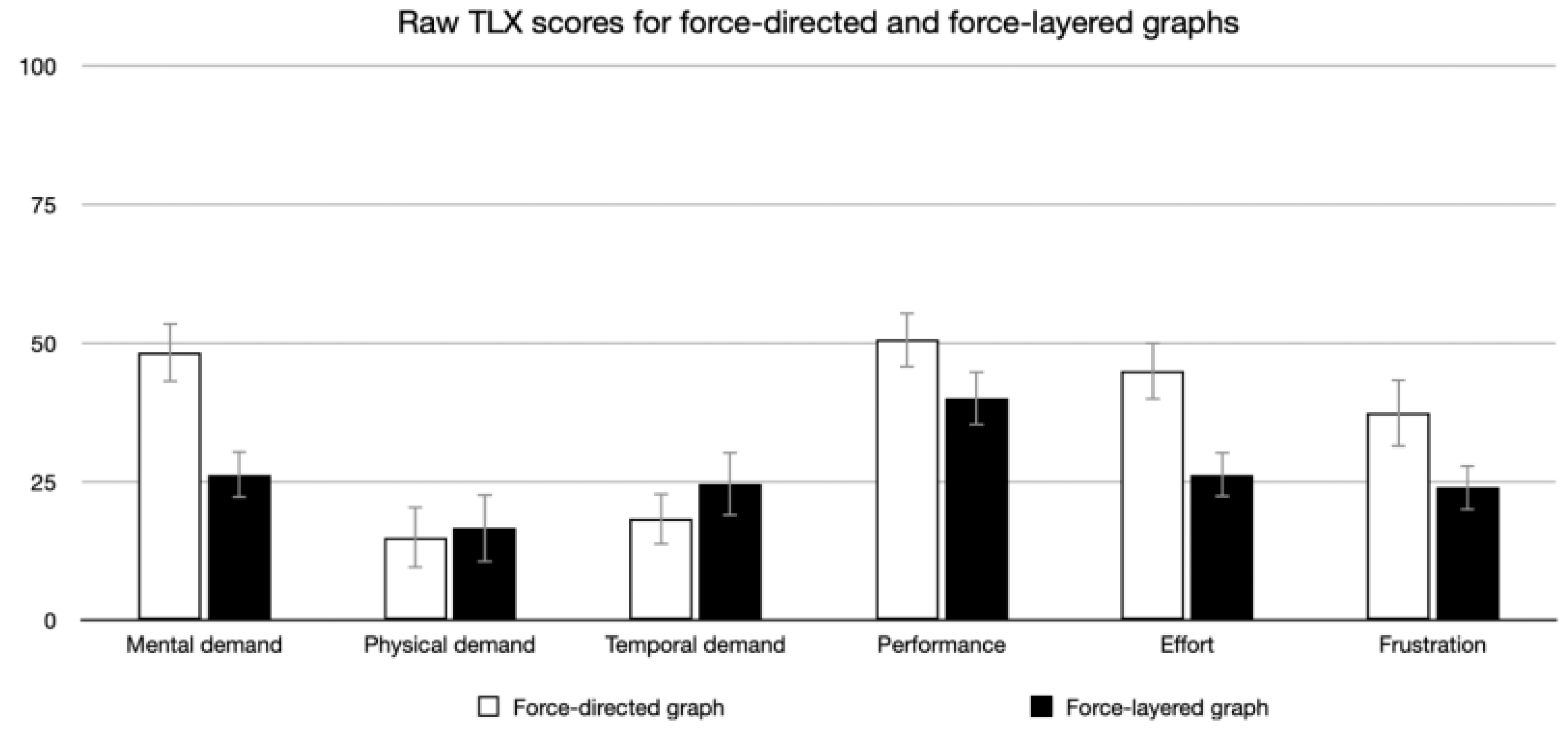

*Figure 6 Results of the raw TLX for the force-layered and force-directed conditions*

[15] Cognitive load is a multidimensional construct representing mental effort – or load imposed on the cognitive system by performing a certain task. Evaluation of this measure has been found to be useful to complement performance-based evaluations of effectiveness in visualization.

[16] In the full version of a NASA-TLX, participants are also asked to weigh these variables depending on their perceived relevance to the task. The final score of a full NASA-TLX is a weighed average of the user score based on the overall weight distribution. However, many researchers report using a *raw* version of the test, where the scores are simply averaged with no consideration of weights. Since little difference in sensitivity has been found between raw and weighed TLX, we decided for a raw-TLX to keep evaluation time short.

## 4.6 User preference

Finally, we performed a preference-based ranking. Participants were asked to subjectively rank the force-directed and the force-layered graphs on a Likert scale on aspects such as perceived suitability to task and data, or visual appeal.

Figure 7 shows the average results of the preference rankings for each category. Globally, participants ranked the force-layered condition higher than the force-directed condition in suitability to both data and tasks. They indicated that the force-layered condition needed less effort to understand than the standard one, and that they had higher confidence in their answers. Finally, the force-layered graphs also scored higher in visual appeal, and overall preference.

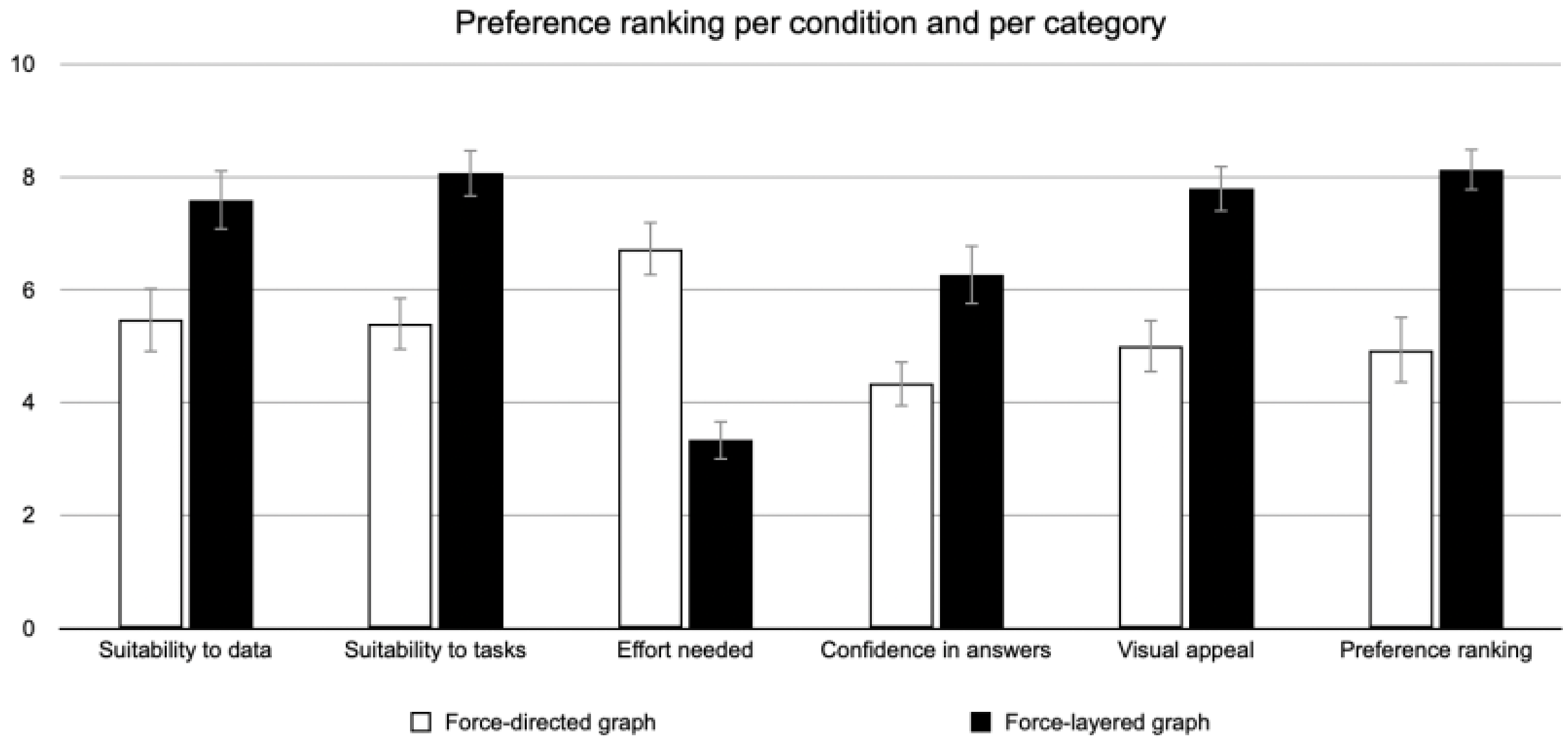

*Figure 7 Results of the preference ranking for the force-layered and force- directed conditions*

# 5. Results & Discussion

Based on a task of mental model elicitation, we found that participants overwhelmingly relied on generational hierarchical structure to draw familial social networks. This can be either because of inherent structure perceived in the data, or association of this type of dataset with traditional ancestry tree representations.

When exploring two different versions of a social network where one is layered hierarchically and the other one isn't, we found that participants expressed more – and deeper – insights in the case of the force-layered visualization. They also engaged more critically with the dataset, for instance by pointing out inconsistencies. The force-directed version, however, drew more comments and descriptions about the visualization features, suggesting that participants engaged more with the *visualization* in the force-directed condition, and with the *data* in the force-layered condition. We also found that the force-layered representation elicited more positive emotions such as curiosity and expressions of enjoyment or interest, that were not as frequent using the force-directed version. Moreover, our results reveal that users perceived the social structure shown in force-layered graphs to be smaller, and less complex than similar structure visualized through a force-directed layout. However, they were more accurate in size estimation using the force-directed layout.

Concerning recall, we found that participants retained more of the significant clusters with the force-layered condition, although they retained more of the bridge nodes after using the force-directed version. In both cases, the bridge nodes were placed in the center of the visualization, so saliency does not explain the difference in these results. We suggest that can be caused by the topology of each graph. In the force-layered version, the ties among each family create two horizontal substructures that become the major ground of the visual, while the dog-bone structure in the force-directed layout represent the major part of the ground figure, and the bridge node is an essential part to its structure. Finally, using a raw NASA-TLX questionnaire and a preference ranking, we found that participants found the force-layered visualization less demanding cognitively. They also judged it to be more suited to the tasks and data, and more visually appealing.

As we analyze these findings in conversation with the literature on visualization in humanistic research, we identify emerging discussion points:

- ***Engagement with data & engagement with visualization.*** In the think-aloud study results, we describe how hierarchical trees seemingly elicit an engagement with the *data*, while force-directed layouts trigger comments about the features and characteristics of the *tool* itself. This subtle distinction has relevant implications, as it suggests that the spatial organization of the network affects the perception of the data representation. Visualized data can therefore be integrated in users' mental models in a way that allows them to analyze and question it; or it can be obscured by the layer of foreign conceptualization that does not allow it to blend with their existing knowledge. As it is, it is critical to encourage representations that allow a connection with the research material itself, and where the visualization features themselves do not become the center of attention for users. We also see this finding as a validating step of the guidelines developed in the Layers of Meaning framework [Lamqaddam, 2020]. In this case, by layering meaning through the structure of network graphs, we find that there is an effect to the level of engagement users have with research data. What seems to be a slight visual difference appears to affect the very focus of users' attention, which in turn reveals the critical aspect of connotative elements and intentional design choices in visualization.

- ***Force-directed network structure as artifact.*** In the analysis of the recall of social structure information, we note that more participants recalled the existence of a bridge node between the two clusters with the force-directed network. We suggest that this is due to the singular topology of the generated network. In truth, what is a particular visual characteristic of a network topology (a bridge node between two tight clusters) is an abstraction of a significant piece of information in the dataset (a common person bringing together two large families). In a way, the fact that the force-directed layout created this artificial shape helped highlight a very important information in the data, which has been better remembered by participants. From this perspective, we can thus defend that force-directed layouts also *create* meaning through the specific topologies they generate. These shapes are often singular, or unique, because they are not a product of intentional design. Rather, then come as the result of computations unique to the dataset, the canvas space and the set parameters. We therefore defend the concept of a force-directed network organization as an artifact in itself. One that holds the potential to carry meaning; to be recognized, analyzed, and categorized.

- ***Defending the importance of challenging mental models.*** We argue throughout this article that one of the strengths of hierarchical tree structures is that they reproduce existing mental models, therefore making it easier to read and grasp information they represent as it fits within scholars' existing knowledge. However, existing models of knowledge are not necessarily correct, accurate, or good to perpetuate. For instance, conventional family trees carry a variety of assumptions about social structures that make them both limited and limiting when representing real-life data. Worse, these representations, when uncritically absorbed, function as prescriptive principles of how communities should exist in society. There is therefore an undeniable larger implication in the discussion about how social network data should be represented. The improved understanding, preference and lower effort that hierarchical trees bring needs to be balanced – or perhaps complemented - by the novel perspectives that force-directed layouts create.

# 6. Implications & Future work

More research is needed to assess the effect of other types of structure in network visualizations. Once these effects are clearly understood, it would be interesting to evaluate how novel graph layout such as our FL algorithm can be introduced to existing visualization packages. It would also be interesting to make them customizable to specific needs outside of family and generational structures. Indeed, an important implication of this work is that any singular off-the-shelf technique cannot be sufficient for a solid support of DAH research. Rather than promote the use of a specific visualization technique over another, we advocate for researchers access to a variety of representation techniques that can accompany different tasks, data types, and expertise levels.

Finally, the generalizability of our discussion to other humanistic fields is an additional avenue for future research. Since this study was mainly focused on an art historical research, our initial results cannot be readily extrapolated to different domains. However, similar challenges of historical social networks exist for humanist researchers across disciplines. Future research is needed to evaluate how hierarchy in historical social network visualizations affects different research fields depending on their unique needs and constraints.

# 7. Conclusion

In this article, we explored the perceptual effect of hierarchy in visualizations of historical social networks. We have found, based on a user-centered approach, that humanist participants had a hierarchical mental model of social networks, indicating that there could be advantages to representing social networks in a similarly hierarchical structure. Our study confirmed this hypothesis, as we found that users reported lower cognitive load, more frequent and deeper insights, as well as a significant preference for the hierarchical representations. Despite these results, we nuance these findings by calling attention to the fact that force-directed graphs can carry meaning in their generated topology, allow users to make conclusions on tightness of networks, clusters, and question the accepted dogmas in terms of community structure. However, because social networks are so critical to certain humanistic research domains, the perceptual benefits to hierarchically structured layouts should not be overlooked. As graph-drawing software is often the most accessible way for researchers to gather quick visual overview, we suggest that these layouts should be more commonly supported as alternatives to allow scholars to have multiple perspectives on their data.

By these findings, we hope to contribute to the necessary critical discussion nuancing the role of network visualization in humanistic research. In the future, we would like to broaden the scope of this study by investigating the effects of hierarchy for other types of historical (social) networks[17]. We believe such an interdisciplinary line of research has the potential to trigger novel questions – for instance about the impact of structure in network visualization of historical data, and the perceptual effects of commonly used design patterns. Only by questioning what we assume to be effective can we (re)design visualization tools that blend in with scholars' experience of their material and practice, and ultimately strengthen the value of visualization as a research tool for humanists.

# Acknowledgements

This research is part of Project Cornelia, a project funded by the University of Leuven (KU Leuven) and the Flemish Fund for Scientific Research-Belgium (FWO-Vlaanderen). We thank our colleagues from AugmentHCI (Department of Computer Science, KU Leuven) and Project Cornelia (Faculty of Arts, KU Leuven) who gave feedback and insight into this study. We are also deeply grateful to the entire NA+DAH team for supporting and encouraging this line of research,

---

[17] Within the Cornelia database, hierarchical relations such as those between masters and apprentices in a workshop are good candidates for such an exploration.